\documentclass[traditabstract]{aa} 

\usepackage{natbib}
\usepackage{amsmath}
\bibpunct{(}{)}{;}{a}{}{,}

\usepackage{graphicx}
\usepackage{txfonts}

\def\ms{\,m\,s$^{-1}$}         
\def\kms{\,km\,s$^{-1}$}       
\def\msol{$M_\odot$}		
\def\rsol{$R_\odot$}		
\def\denssol{$\rho_\odot$}	
\def\mstar{$M_*$}		
\def\rstar{$R_*$}		
\def\densstar{$\rho_*$}		
\def\mplanet{$M_{\rm P}$}	
\def\rplanet{$R_{\rm P}$}	
\def\densplanet{$\rho_{\rm P}$}
\def\mjup{$M_{\rm Jup}$}	
\def\rjup{$R_{\rm Jup}$}	
\def\densjup{$\rho_{\rm Jup}$}	
\def\teql{$T_{\rm eql}$}
\def\teff{$T_{\rm eff}$}
\def\feh{[Fe/H]}
\def\logg{$\log g_*$}
\def\vsini{$v \sin I_*$}
\def\mictrb{$\xi_{\rm t}$}
\def\mactrb{$v_{\rm mac}$}

\def\halpha{$H_\alpha$}
\def\kms{km\, s$^{-1}$}

\def\ecos{$e \cos \omega$}
\def\esin{$e \sin \omega$}
\def\secos{$\sqrt{e} \cos \omega$}
\def\sesin{$\sqrt{e} \sin \omega$}
\def\svsicos{$\sqrt{v \sin I} \cos \lambda$}
\def\svsisin{$\sqrt{v \sin I} \sin \lambda$}
\def\chisq{$\chi^2$}
\def\deg{$^\circ$}

\newcommand{\leftcell}[1]{\multicolumn{1}{l}{#1}}

\begin{document}

   \title{WASP-20b and WASP-28b: a hot Saturn and a hot Jupiter in near-aligned orbits around solar-type stars\thanks{
Based on observations made with: the WASP-South (South Africa) and SuperWASP-North (La Palma) 
photometric survey instruments; the C2 and EulerCam cameras and the CORALIE spectrograph, all mounted on the 1.2-m Euler-Swiss telescope (La Silla); the HARPS spectrograph on the ESO 3.6-m telescope (La Silla) under programs  
072.C-0488, 082.C-0608, 084.C-0185 and 085.C-0393; and LCOGT's Faulkes Telescope North (Maui) and 
Faulkes Telescope South (Siding Spring).}
}


   \titlerunning{WASP-20b and WASP-28b}

\author{D.~R.~Anderson
        \inst{1}
	\and
A.~Collier~Cameron
        \inst{2}
	\and 
C.~Hellier
        \inst{1}
	\and 
M.~Lendl
        \inst{3}
	\and 
T.~A.~Lister
        \inst{4}
	\and 
P.~F.~L.~Maxted
        \inst{1}
	\and 
D.~Queloz
        \inst{3,5}
	\and 
B.~Smalley
        \inst{1}
	\and 
A.~M.~Smith
        \inst{1,6}
	\and 
A.~H.~M.~J.~Triaud
        \inst{3,7}
	\and
D.~J.~A.~Brown
		\inst{8,9}
	\and
M.~Gillon
        \inst{10}
	\and 
M.~Neveu-VanMalle
        \inst{3}
	\and 
F.~Pepe
        \inst{3}
	\and 
D.~Pollacco
        \inst{8}
	\and 
D.~S\'egransan
        \inst{3}
	\and 
S.~Udry
        \inst{3}
	\and
R.~G.~West
        \inst{8}
	\and
P.~J.~Wheatley
        \inst{8}
}

\institute{Astrophysics Group, Keele University, Staffordshire ST5 5BG, UK\\
           \email{d.r.anderson@keele.ac.uk}
	   \and
			SUPA, School of Physics and Astronomy, University of St. Andrews, 
           North Haugh, Fife KY16 9SS, UK
		\and
			Observatoire de Gen\`eve, Universit\'e de Gen\`eve, 51 Chemin 
           des Maillettes, 1290 Sauverny, Switzerland
		\and
			Las Cumbres Observatory Global Telescope Network, 6740 Cortona Dr. Suite 102, Goleta, CA 93117, USA
		\and
			Cavendish Laboratory, J J Thomson Avenue, Cambridge, CB3 0HE, UK
		\and
			N. Copernicus Astronomical Centre, Polish Academy of Sciences, 
			Bartycka 18, 00-716, Warsaw, Poland
		\and
			Department of Physics, and Kavli Institute for Astrophysics and Space Research,
			MIT, Cambridge, MA 02139, USA
		\and
			Department of Physics, University of Warwick, Coventry CV4 7AL, UK
		\and
			Astrophysics Research Centre, School of Mathematics \& Physics, Queen's University, University Road, Belfast BT7 1NN
		\and
			Institut d'Astrophysique et de G\'eophysique,  Universit\'e de 
           Li\`ege,  All\'ee du 6 Ao\^ut, 17,  Bat.  B5C, Li\`ege 1, Belgium
}

   \date{Received February 6, 2014 / accepted Month DD, YEAR}

 
\abstract{We report the discovery of the planets WASP-20b and WASP-28b along with measurements of their sky-projected orbital obliquities. 
WASP-20b is an inflated, Saturn-mass planet (0.31\,\mjup; 1.46\,\rjup) in a 4.9-day, near-aligned ($\lambda = 8.1 \pm 3.6 \degr$) orbit around CD-24~102 ($V$=10.7; F9). 
WASP-28b is an inflated, Jupiter-mass planet (0.91\,\mjup; 1.21\,\rjup) in a 3.4-day, near-aligned ($\lambda = 8 \pm 18 \degr$) orbit around a $V$=12, F8 star. 
As intermediate-mass planets in short orbits around aged, cool stars ($7^{+2}_{-1}$ Gyr for WASP-20 and $5^{+3}_{-2}$ Gyr for WASP-28; both with \teff\ $<$ 6250\,K), their orbital alignment is consistent with the hypothesis that close-in giant planets are scattered into eccentric orbits with random alignments, which are then circularised and aligned with their stars' spins via tidal dissipation. 
}


   \maketitle
%

\section{Introduction}

Planets that transit relatively bright host stars ($V$$<$13) are proving a rich 
source of information for the nascent field of exoplanetology. 
To date, the main discoverers of these systems are the ground-based transit 
surveys HATNet and SuperWASP and the space mission Kepler 
\citep{2004PASP..116..266B, 2006PASP..118.1407P, 2010Sci...327..977B}. 

One parameter that, uniquely, we can determine for transiting planets is 
obliquity ($\Psi$), the angle between a star's rotation axis and a planet's 
orbital axis. 
We do this by taking spectra of a star during transit: 
as the planet obscures a portion of the rotating star it causes a distortion of  
the observed stellar line profile, which manifests as an anomalous 
radial-velocity (RV) signature known as the Rossiter-McLaughlin (RM) effect 
\citep{1924ApJ....60...15R,1924ApJ....60...22M}.
The shape of the RM effect is sensitive to the path a planet takes across the 
disc of a star relative to the stellar spin axis. 
If we have a constraint on the inclination of the stellar 
spin axis relative to the sky plane ($I_*$), then we can determine the true 
obliquity ($\Psi$), otherwise we can only determine the sky-projected obliquity 
($\lambda$). The two are related by:

$\cos \Psi = \cos I_* \cos i_{\rm P} + \sin I_* \sin i_{\rm P} \cos \lambda$

\noindent where $i_{\rm P}$ is the inclination of the orbital axis to the sky plane.

The obliquity of a short-period, giant planet may be indicative 
of the manner in which it arrived in its current orbit from farther out, where 
it presumably formed. 
As the angular momenta of a star and its planet-forming disc both derive from 
that of their parent molecular cloud, stellar spin and planetary orbital axes 
are expected to be, at least initially, aligned. 
Migration via tidal interaction with the gas disc is expected to preserve  
initial spin-orbit alignment \citep{1996Natur.380..606L,2009ApJ...705.1575M}, 
but almost half (33 of 74) of the orbits so far measured are misaligned and 
approximately 10 of those are retrograde.\footnote{Ren\'e Heller maintains a list of 
measurements and references at 
\url{http://www.physics.mcmaster.ca/~rheller/content/main_HRM.html}}

These results are consistent with the hypothesis that 
some or all close-in giant planets 
arrive in their orbits by planet-planet and/or star-planet scattering, which can drive planets into eccentric, 
misaligned orbits, and tidal friction, which circularises, shortens and aligns orbits 
(see the following empirical-based papers: \citealt{2010A&A...524A..25T,2010ApJ...718L.145W,2011Natur.473..187N,2012ApJ...757...18A}; 
also see the following model-based papers: 
\citealt{2007ApJ...669.1298F,2008ApJ...678..498N,2010ApJ...725.1995M,
2011Natur.473..187N}). 

Systems with short tidal timescales (those with short scaled orbital major semi-axes, 
$a$/\rstar, and high planet-to-star mass ratios) tend to be aligned 
(\citealt{2012ApJ...757...18A} and references therein). 
A broad range of obliquities is observed for stars with \teff$>$6250\,K, for which 
tidal realignment processes may be inefficient due to the absence of a 
substantial convective envelope \citep{2010ApJ...718L.145W,2010ApJ...719..602S}. 
Limiting focus to stars more massive than 1.2\,\msol, for which age determinations 
are more reliable, \citet{2011A&A...534L...6T} noted that systems older than 2.5~Gyr 
tend to be aligned; 
this could be indicative of the time required for orbital alignment or of the timescale 
over which hotter stars develop a substantial convective envelope as they evolve. 

A major hurdle to overcome for any hypothesis involving realignment is tidal 
dissipation seems to cause both orbital decay and realignment on similar timescales 
\citep{2009MNRAS.395.2268B}. 
Alternative hypotheses suggest that misalignments arise via reorientations of either the 
disc or the stellar spin and that migration then proceeds via planet-disc interactions 
(e.g. \citealt{2010MNRAS.401.1505B,2011MNRAS.412.2790L,2012ApJ...758L...6R}).
However, observations of debris discs well aligned with their stellar equators suggest that 
tilting of the star or the disc rarely occurs \citep{2014MNRAS.438L..31G}.

Here we present the discovery and obliquity determinations of the two new transiting 
planets WASP-20b (CD-24~102) and WASP-28b (2MASS~J23342787$-$0134482), 
and interpret the results under the hypothesis of migration via scattering and tidal 
dissipation. 

\section{Observations}
\begin{figure}
\includegraphics[width=84mm]{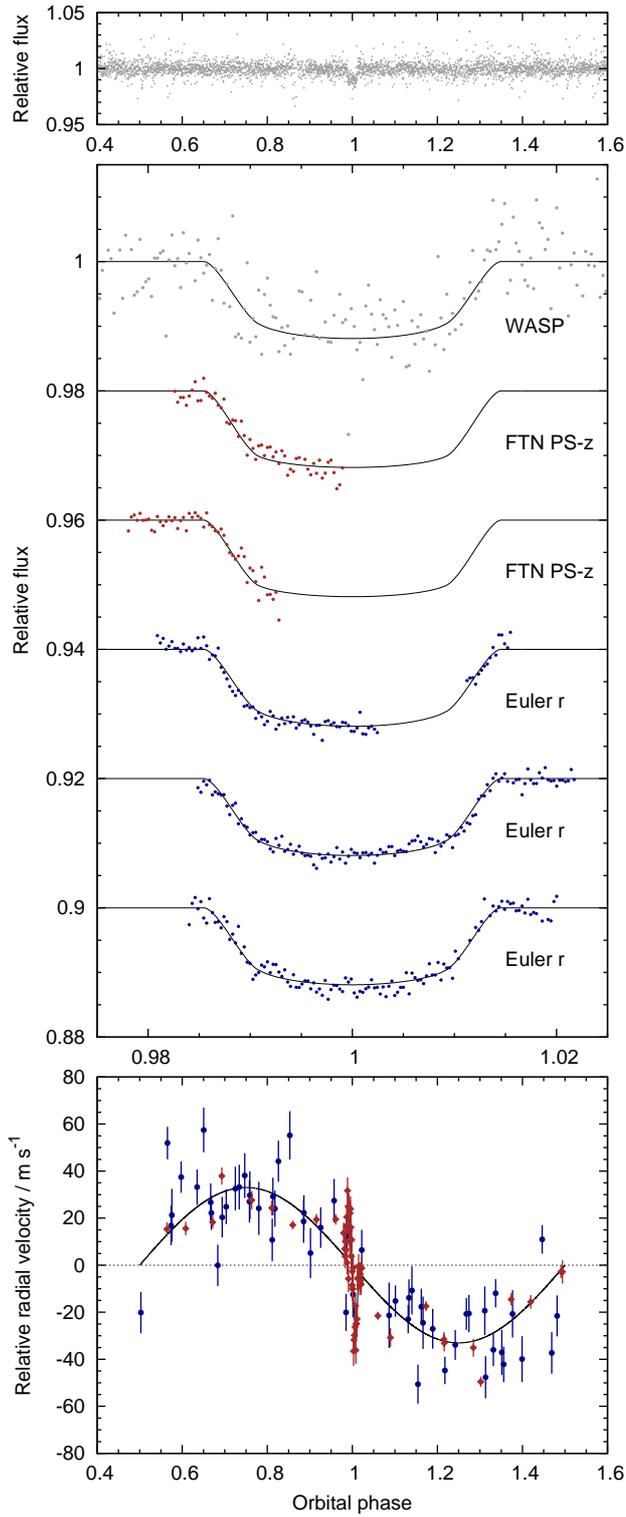}
\caption{WASP-20b discovery data. 
{\it Top panel}: WASP lightcurve folded on the transit ephemeris. 
{\it Middle panel}: Transit lightcurves from facilities as labelled, offset for clarity and binned with a bin width of two minutes. 
The best-fitting transit model is superimposed. 
{\it Bottom panel}: The radial velocities (CORALIE in blue, HARPS in brown) with the best-fitting circular Keplerian orbit model and the RM effect model. 
\label{fig:w20-rv-phot}}
\end{figure}

\begin{figure}
\includegraphics[width=84mm]{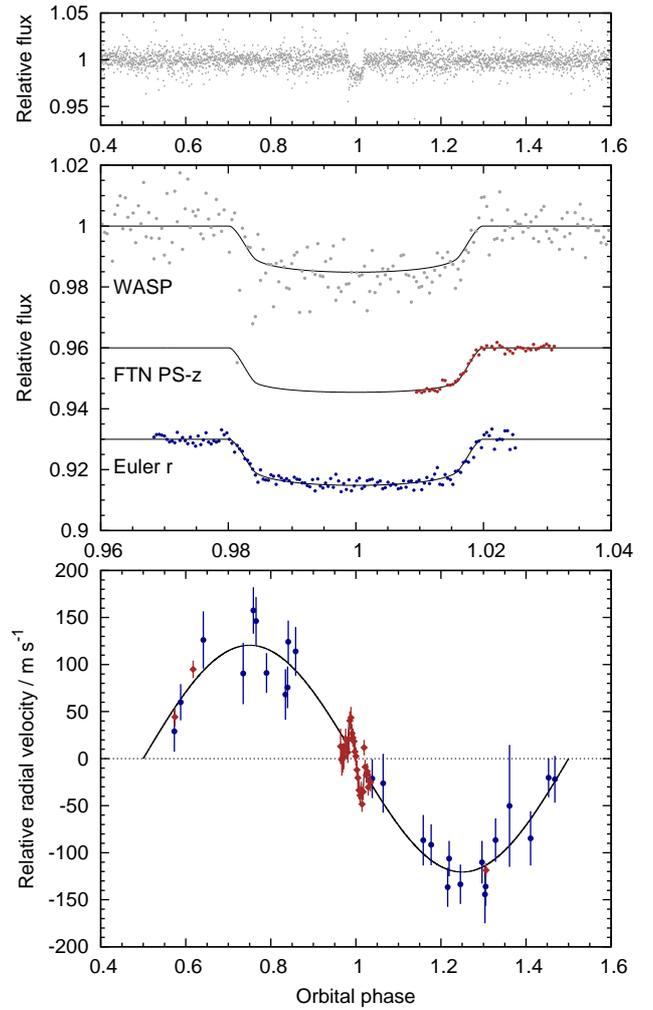}
\caption{WASP-28b discovery data. 
Caption as for Figure~\ref{fig:w20-rv-phot}. 
\label{fig:w28-rv-phot}}
\end{figure}

We provide a summary of observations in Table~\ref{tab:obs}. 
The WASP (Wide Angle Search for Planets) photometric survey 
\citep{2006PASP..118.1407P} monitors bright stars ($V$ = 9.5--13.5) 
using two eight-camera arrays, each with a field of view of 450 deg$^2$. 
Each array observes up to eight pointings per night with a cadence of 
5--10 min, and each pointing is followed for around five months per season. 
The WASP-South station \citep{2011EPJWC..1101004H} is hosted by 
the South African Astronomical Observatory and the SuperWASP-North station 
\citep{2011EPJWC..1101003F} is hosted by 
the Isaac Newton Group at the Observatorio del Roque de Los Muchachos on 
La Palma. 
The WASP data were processed and searched for transit signals as described in 
\citet{2006MNRAS.373..799C} and the candidate selection process was performed 
as described in \citet{2007MNRAS.380.1230C}.
We observed periodic dimmings in the WASP lightcurves of WASP-20 and WASP-28 
with periods of 4.8996\,d and 3.4088\,d, respectively (see the top panels of 
Figures~\ref{fig:w20-rv-phot} and \ref{fig:w28-rv-phot}). 
We searched the WASP lightcurves for modulation as could be caused by magnetic 
activity and  stellar rotation \citep{2011PASP..123..547M}. 
We did not detect any significant, repeated signals and we find that any 
modulation must be below the 2-mmag level. 

\begin{figure}
\includegraphics[width=84mm]{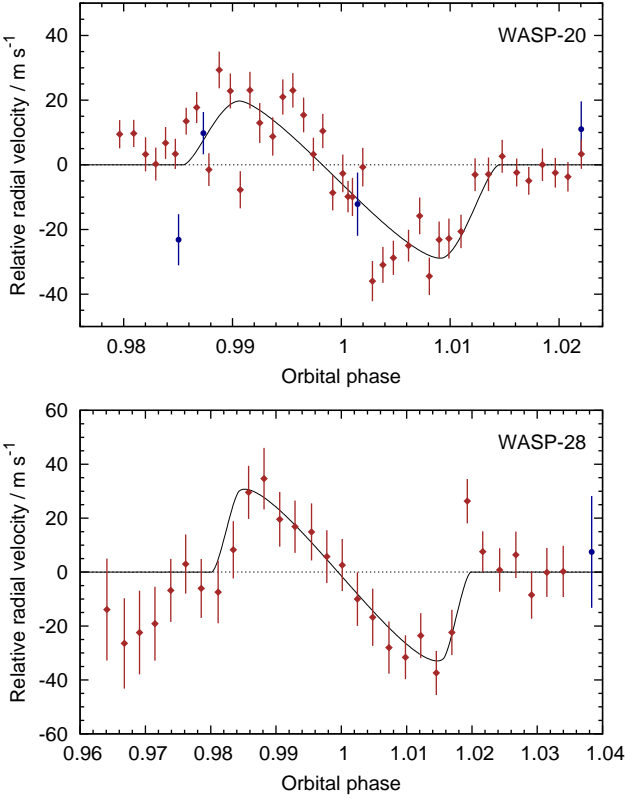}
\caption{The Rossiter-McLaughlin effects, or spectroscopic transits, of 
WASP-20b and WASP-28b. The CORALIE RVs are shown in blue and the HARPS RVs are shown in brown; 
the Keplerian orbits have been subtracted. 
For each planet, the observed apparent redshift, followed by an apparent 
blueshift of similar amplitude, indicate the orbit to be prograde and, 
probably, aligned with the stellar rotation. 
The alignment of the orbit of WASP-28b is less certain due to its low 
impact parameter. 
The lower precision of the earlier WASP-28 RVs resulted because the sequence 
started at an airmass of 1.7. 
 \label{fig:rm}}
\end{figure}

We obtained 56 spectra of WASP-20 and 26 spectra of WASP-28 with the CORALIE spectrograph mounted on the Euler-Swiss 1.2-m telescope 
\citep{1996A&AS..119..373B,2000A&A...354...99Q}. 
We obtained a further 63 spectra of WASP-20 with the HARPS spectrograph mounted on the 3.6-m ESO telescope \citep{2003Msngr.114...20M}, including a sequence of 43 spectra taken through the transit of the night of 2009 October 22. 
We obtained a further 33 spectra of WASP-28 with HARPS, including 30 spectra taken through the transit of the night of 2010 August 17. 
Radial-velocity (RV) measurements were computed by weighted cross-correlation
\citep{1996A&AS..119..373B,2005Msngr.120...22P} with a numerical G2-spectral
template (Table~\ref{rvs}). 
We detected RV variations with periods similar to those found from the WASP photometry and with semi-amplitudes consistent with planetary-mass companions. 
We plot the RVs, phased on the transit ephemerides, in the third panel of 
Figures~\ref{fig:w20-rv-phot} and \ref{fig:w28-rv-phot} and we highlight 
the RM effects in Figure~\ref{fig:rm}

\begin{figure}[h!]
\includegraphics[width=84mm]{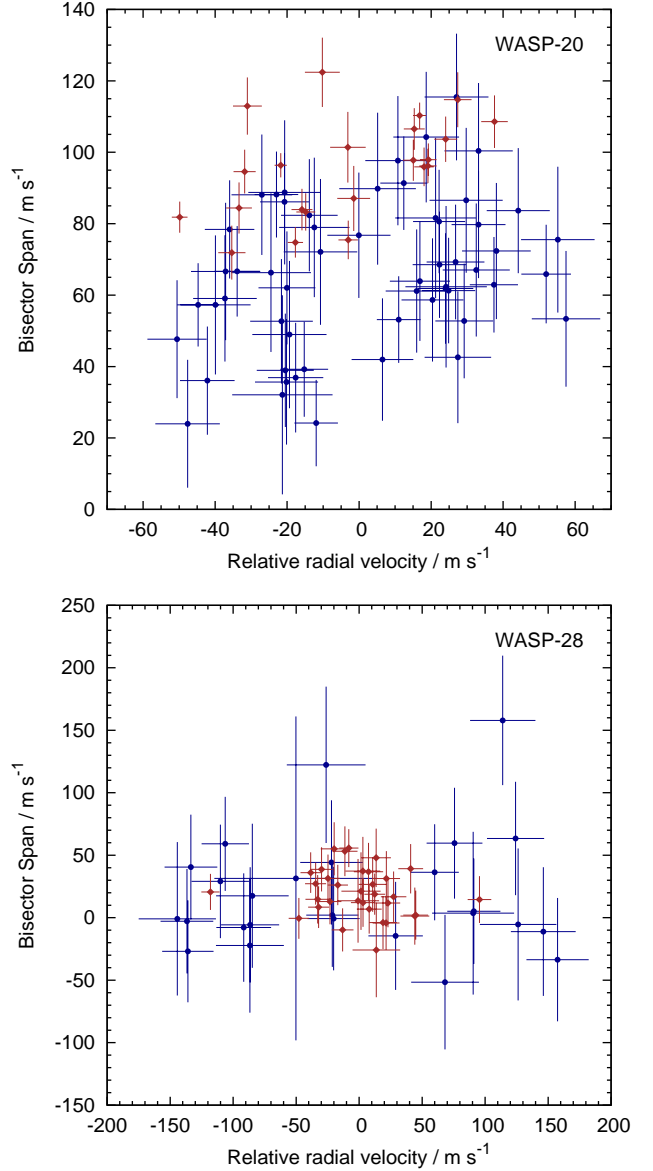}
\caption{The lack of any significant correlation between bisector span and 
radial velocity excludes transit mimics and supports our conclusion 
that each system contains a transiting planet. 
\label{fig:bis}}
\end{figure}

For each star, we tested the hypothesis that the RV variations are due to
spectral-line distortions caused by a blended eclipsing binary or starspots
by performing a line-bisector analysis of the
cross-correlation functions \citep{2001A&A...379..279Q}.
The lack of any significant correlation between bisector span and RV supports 
our conclusion that the periodic dimming and RV variation of each system are 
caused by a transiting planet (Figure~\ref{fig:bis}). 
This is further supported by our observation of the RM effect of each system: 
the \vsini\ values from our fits to the RM effects are consistent with the 
values we obtain from spectral line broadening (see Sections \ref{sec:stars} 
and \ref{sec:mcmc}).

We performed follow-up transit observations to refine the systems' 
parameters using the 1.2-m Swiss Euler telescope \citep{2012A&A...544A..72L}, 
and LCOGT's 2.0-m Faulkes Telescopes North and South 
\citep{2013PASP..125.1031B}. Some transit lightcurves were affected by 
technical problems or poor weather, as indicated in Table~\ref{tab:obs}. 

\begin{table*}
\centering
\caption{Summary of observations}
\label{tab:obs}
\begin{tabular}{lcrcll}
\hline
\leftcell{Facility} & \leftcell{Date} & \leftcell{$N_{\rm obs}$} & \leftcell{$T_{\rm exp}$} & \leftcell{Filter} & \leftcell{Issue} \\
         &      &      & \leftcell{(s)}           &        \\
\hline
{\bf WASP-20:}\\
WASP-South	    & 2006 May--2007 Nov	& 9\,600& 30 & Broad (400--700 nm)	& --- \\
Euler/CORALIE	& 2008 Jul--2013 Oct	& 56	& 1800		& Spectroscopy  & --- \\
ESO-3.6m/HARPS	& 2008 Aug--2011 Sep	& 20	& 600--1800	& Spectroscopy  & --- \\
FTS/Spectral	& 2008 Oct 25			& 151	& 40 & $z'$					& autoguider\\
ESO-3.6m/HARPS	& 2009 Oct 22			& 43	& 300--400/1200	& Spectroscopy  & --- \\
FTN/Merope		& 2011 Aug 20			&  70	& 70 & $z'$					& rotator\\
Euler/EulerCam	& 2011 Aug 29			& 237	& 20 & Gunn-$r$ 			& cloud\\
Euler/EulerCam	& 2011 Sep 03			& 156	& 70 & Gunn-$r$ 			& --- \\
Euler/EulerCam	& 2011 Sep 08			& 201	& 30 & Gunn-$r$ 			& --- \\
{\bf WASP-28}\\
WASP-South	    & 2008 Jun--2009 Nov	& 10\,100 & 30 & Broad (400--700 nm)& --- \\
SuperWASP-North & 2008 Aug--2010 Sep	&  6\,600 & 30 & Broad (400--700 nm)& --- \\
Euler/CORALIE	& 2009 Jun--2012 Dec	&  26	  & 1800 & Spectroscopy		& --- \\
FTN/Merope		& 2009 Oct 21			& 227	  & 40 & $z'$				& unknown\\
ESO-3.6m/HARPS	& 2010 Aug 17--19		& 33	  & 600/1200 & Spectroscopy & --- \\
Euler/C2		& 2010 Sep 03			& 217	  & 25 & Gunn-$r$ 			& --- \\
\hline
\end{tabular}
\end{table*}

\begin{table}
\caption{Radial-velocity measurements} 
\label{rvs} 
\resizebox{0.48\textwidth}{!}{%
\begin{tabular}{llrrrrr} 
\hline 
\leftcell{Star} & Spect. & \leftcell{BJD(UTC)} & \leftcell{RV} & \leftcell{$\sigma$$_{\rm RV}$} & \leftcell{BS}\\ 
    & & \leftcell{$-$2450000}&    &                     &   \\
    & & \leftcell{(day)}     & \leftcell{(km s$^{-1}$)} & \leftcell{(km s$^{-1}$)} & \leftcell{(km s$^{-1}$)}\\ 
\hline
W-20 & COR & 4655.88265 &  1.3411 & 0.0074 & 0.0613 \\
W-20 & COR & 4658.86065 &  1.2969 & 0.0103 & 0.0490 \\
\ldots\\
W-28 & HAR & 5428.71664 & 24.2674 & 0.0098 & 0.0021 \\
W-28 & HAR & 5428.86297 & 24.3180 & 0.0094 & 0.0145 \\
\hline
\end{tabular}}
The uncertainties are the formal errors (i.e. with no added jitter).\\  
The uncertainties on the bisector spans (BS) are 2\,$\sigma_{\rm RV}$.\\
This table is available in its entirety via the CDS. 
\end{table}

\begin{table} 
\caption{Follow-up photometry} 
\label{tab:phot} 
\begin{tabular*}{0.48\textwidth}{lllrrr} 
\hline 
\leftcell{Set} & \leftcell{Star} & \leftcell{Tel.} & \leftcell{BJD(UTC)} & \leftcell{Flux, $F$} & \leftcell{$\sigma_{F}$}\\ 
    &      &      & \leftcell{$-$2450000} &        &                 \\
& & & \leftcell{(day)} & & \\
\hline
1	& W-20 & FTN & 4765.04209 & 0.99958 & 0.00179 \\
1	& W-20 & FTN & 4765.04283 & 1.00022 & 0.00163 \\
\ldots \\
7	& W-28 & Euler & 5443.88762 & 0.9971~~ & 0.0027~~ \\
7	& W-28 & Euler & 5443.88852 & 1.0006~~ & 0.0027~~ \\
\hline
\end{tabular*}
\begin{flushleft}
The flux values are differential and normalised to the out-of-transit levels. 
\newline The uncertainties are the formal errors (i.e. they have not been rescaled).
\newline This table is available in its entirety via the CDS. 
\end{flushleft}
\end{table}

\clearpage

\clearpage

\section{Stellar parameters from spectra}
\label{sec:stars}
The 9 individual HARPS spectra from 2008 of WASP-20 were co-added and a total of 18 individual CORALIE spectra were co-added to produce one spectrum per star 
with typical S/N of 150:1 (WASP-20) and 70:1 (WASP-28).
The standard pipeline reduction products were used in the analysis.

The analysis was performed using the methods given in Gillon et al. (2009).
The \halpha\ line was used to determine the
effective temperature (\teff), while the Na {\sc i} D and Mg {\sc i} b lines
were used as surface gravity (\logg) diagnostics. The parameters obtained from
the analysis are listed in Table~\ref{tab:stars}. The elemental abundances
were determined from equivalent width measurements of several clean and
unblended lines. A value for microturbulence (\mictrb) was determined from
Fe~{\sc i} using Magain's (1984) method. The quoted error estimates include
that given by the uncertainties in \teff, \logg\ and \mictrb, as well as the
scatter due to measurement and atomic data uncertainties.

We assumed values for macroturbulence (\mactrb) of 4.5 $\pm$ 0.3 \kms\ and 4.7 $\pm$ 0.3 \kms\ 
for WASP-20 and WASP-28, respectively. 
These were based on the tabulation by Gray (2008) and 
the instrumental FWHMs of 0.065\AA\ (HARPS) and 0.11 $\pm$ 0.01 \AA\ (CORALIE), as 
determined from the telluric lines around 6300\AA. 
We determined the projected stellar rotation velocity (\vsini) by fitting the
profiles of several unblended Fe~{\sc i} lines. We determined the \vsini\ of WASP-20 
from the 62 HARPS spectra obtained up 2010 and the \vsini\ 
of WASP-28 from all 33 available HARPS spectra.  

\begin{table}
\caption{Stellar parameters from the spectra}
\begin{tabular*}{0.5\textwidth}{lcc}
\hline
\leftcell{Parameter}	& WASP-20	& WASP-28 \\
\hline
Star			& CD-24~102			& \small{2MASS~J23342787$-$0134482} \\
Constellation	& Cetus				& Pisces \\
R.A. (J2000)	& $\rm  00^{h} 20^{m} 38\fs53$ & $\rm  23^{h} 34^{m} 27\fs87$ \\
Dec. (J2000)	& $\rm -23\degr 56\arcmin 08\farcs6$ & $\rm -01\degr 34\arcmin 48\farcs2$ \\
$B$				& 11.17 & 12.50 \\
$V$   			& 10.68 & 12.03 \\
$J$   			& 9.70  & 11.08 \\
$H$   			& 9.42  & 10.76 \\
$K$				& 9.39  & 10.73 \\
\teff\ 			& 6000 $\pm$ 100 K		& 6100 $\pm$ 150 K\\
\logg			& 4.40 $\pm$ 0.15	& 4.5 $\pm$ 0.2 \\
\mictrb\ 		& 1.2 $\pm$ 0.1	\kms& 1.2 $\pm$ 0.1 \kms\\
\vsini\ 		& 3.5 $\pm$ 0.5 \kms& 3.1 $\pm$ 0.6 \kms\\
{[Fe/H]}   		& $-$0.01 $\pm$ 0.06& $-$0.29 $\pm$ 0.10 \\
{[Si/H]}		& +0.03 $\pm$ 0.09	& $-$0.22 $\pm$ 0.10 \\
{[Ca/H]}		& +0.09 $\pm$ 0.11	& $-$0.20 $\pm$ 0.12 \\
{[Sc/H]}		& +0.03 $\pm$ 0.06	& \ldots \\
{[Ti/H]}		& +0.09 $\pm$ 0.11	& $-$0.21 $\pm$ 0.07 \\
{[V/H]}			&  +0.09 $\pm$ 0.07	& \ldots \\
{[Cr/H]}		&$-$0.04 $\pm$ 0.06	& \ldots \\
{[Mn/H]}		&$-$0.01 $\pm$ 0.08	& \ldots \\
{[Co/H]}		&$-$0.02 $\pm$ 0.08	& \ldots \\
{[Ni/H]}		&\phantom{+}0.00 $\pm$ 0.06 & $-$0.28 $\pm$ 0.08 \\
log A(Li)		& \phantom{+}2.40 $\pm$ 0.10	& \phantom{+}2.52 $\pm$ 0.12 \\
Spectral type	& F9				& F8 \\
Age 			& $7^{+2}_{-1}$	Gyr	& $5^{+3}_{-2}$ Gyr \\
Distance		& $210 \pm 20$ pc	& $410 \pm 70$ pc\\
\hline
\end{tabular*}
\label{tab:stars}
\newline {\bf Notes:} We determined \vsini\ from different subsets of spectra than the other parameters, as described in the text.
The magnitudes are from the NOMAD catalogue \citep{2004AAS...205.4815Z}. 
The spectral types are estimated from \teff\ using Table B.1 in \citet{2008oasp.book.....G}.
\end{table}

\section{System parameters from the RV and transit data}
\label{sec:mcmc}
We determined the parameters of each system from a simultaneous fit to the lightcurve and radial-velocity data. 
The fit was performed using the current version of the 
Markov-chain Monte Carlo (MCMC) code described by \citet{2007MNRAS.380.1230C} 
and \citet{2008MNRAS.385.1576P}. 

The transit lightcurves are modelled using the formulation of 
\citet{2002ApJ...580L.171M} with the assumption that the planet is much smaller 
than the star. 
Limb-darkening was accounted for using a four-coefficient, nonlinear 
limb-darkening model, using coefficients appropriate to the passbands from the 
tabulations of \citet{2000A&A...363.1081C, 2004A&A...428.1001C}. 
The coefficients are interpolated once using the values of \logg\ and \feh\ in 
Table~\ref{tab:stars}, but are interpolated at each MCMC step using the 
latest value of \teff. The coefficient values corresponding to the best-fitting 
value of \teff\ are given in Table~\ref{tab:ld}.
The transit lightcurve is parameterised by the epoch of mid-transit 
$T_{\rm 0}$, the orbital period $P$, the planet-to-star area ratio 
(\rplanet/\rstar)$^2$, the approximate duration of the transit from initial to 
final contact $T_{\rm 14}$, and the impact parameter $b = a \cos i_{\rm P}/R_{\rm *}$ 
(the distance, in fractional stellar radii, of the transit chord from the 
star's centre in the case of a circular orbit), where $a$ is the semimajor axis 
and $i_{\rm P}$ is the inclination of the orbital plane with respect to the sky plane. 

\begin{table*}
\centering
\caption{Limb-darkening coefficients} 
\label{tab:ld} 
\begin{tabular}{llllllll}
\hline
\leftcell{Planet}		& \leftcell{Instrument}		& \leftcell{Observation bands}		& \leftcell{Claret band}			& \leftcell{$a_1$}		& \leftcell{$a_2$}		& \leftcell{$a_3$}		& \leftcell{$a_4$}	\\
\hline
WASP-20		& WASP / EulerCam	& Broad (400--700 nm) / Gunn $r$& Cousins $R$	& 0.577 & $-$0.042 & 0.484 & $-$0.290 \\
WASP-20		& FTN / FTS			& Sloan $z'$					& Sloan $z'$	& 0.652 & $-$0.345 & 0.635 & $-$0.325 \\
WASP-28		& WASP / EulerCam	& Broad (400--700 nm) / Gunn $r$& Cousins $R$	& 0.428 & 0.456 & $-$0.175 & $-$0.023 \\
WASP-28		& FTN			& Sloan $z'$					& Sloan $z'$	& 0.516 & 0.036 & 0.160 & $-$0.135 \\
\hline
\end{tabular}
\end{table*}

The eccentric Keplerian radial-velocity orbit is parameterised by the stellar 
reflex velocity semi-amplitude $K_{\rm 1}$, the systemic velocity $\gamma$, 
an instrumental offset between the HARPS and CORALIE spectrographs
$\Delta \gamma_{\rm HARPS}$, and 
\secos\ and \sesin\,  where $e$ is orbital 
eccentricity and $\omega$ is the argument of periastron. 
We use \secos\ and \sesin\ as they impose a uniform prior on $e$, whereas 
the jump parameters we used previously, \ecos\ and \esin, impose a linear prior 
that biases $e$ toward higher values \citep{2011ApJ...726L..19A}.
The RM effect was modelled using the formulation of \citet{2006ApJ...650..408G} 
and, for similar reasons, is parameterised by \svsicos\ and \svsisin.

The linear scale of the system depends on the orbital separation $a$ which, 
through Kepler's third law, depends on the stellar mass \mstar. 
At each step in the Markov chain, the latest values of \densstar, \teff\ and 
\feh\ are input in to the empirical mass calibration of 
\citet{2010A&A...516A..33E} (as based upon \citealt{2010A&ARv..18...67T} and 
as updated by \citealt{2011MNRAS.417.2166S}) to obtain \mstar.
The shapes of the transit lightcurves and the radial-velocity curve constrain 
stellar density \densstar\ \citep{2003ApJ...585.1038S}, which combines with \mstar\ to give 
the stellar radius \rstar.
The stellar effective temperature \teff\ and metallicity \feh\ are proposal parameters constrained by Gaussian priors with 
mean values and variances derived directly from the stellar spectra 
(see Section~\ref{sec:stars}). 

As the planet-to-star area ratio is determined from the measured transit depth, 
the planet radius \rplanet\ follows from \rstar. The planet mass \mplanet\ is calculated from 
the measured value of $K_{\rm 1}$ and the value of \mstar; the planetary density 
\densplanet\ and surface gravity $\log g_{\rm P}$ then follow. 
We  calculate the planetary equilibrium temperature \teql, assuming zero 
albedo and efficient redistribution of heat from the planet's 
presumed permanent day side to its night side. 
We also calculate the durations of transit ingress ($T_{\rm 12}$) and egress 
($T_{\rm 34}$). 

At each step in the MCMC procedure, model transit lightcurves and radial 
velocity curves are computed from the proposal parameter values, which are 
perturbed from the previous values by a small, random amount. The \chisq\ 
statistic is used to judge the goodness of fit of these models to the data and the decision as to whether to accept a step is made via the Metropolis-Hastings rule \citep{2007MNRAS.380.1230C}:
a step is accepted if \chisq\ is lower than for the previous step and a step 
with higher \chisq\ is accepted with a probability 
proportional to $\exp(-\Delta \chi^2/2)$. 
This gives the procedure some robustness against local minima and results in a thorough exploration of the parameter space around the best-fitting solution. 
To give proper weighting to each photometric data set, the uncertainties were 
scaled at the start of the MCMC so as to obtain a photometric reduced-\chisq\ 
of unity. 
To obtain a spectroscopic reduced-$\chi^2$ of unity we added `jitter' terms in quadrature 
to the formal RV errors of WASP-20: 12\,\ms\ for the CORALIE RVs and 7\,\ms\ for the HARPS RVs; 
no jitter was required for WASP-28.

For both WASP-20b and WASP-28b we find, using the $F$-test approach of \citet{1971AJ.....76..544L}, that the improvement in the fit to the RV data resulting from the use of an eccentric orbit model is small and is consistent with the underlying orbit being circular. 
We thus adopt circular orbits, which \citet{2012MNRAS.422.1988A} suggest is the prudent choice for short-period,$\sim$Jupiter-mass planets in the absence of evidence to the contrary.
We find 2-$\sigma$ upper limits on $e$ of 0.11 and 0.14 for WASP-20b and WASP-28b, respectively.

Due to the low impact parameter of WASP-28b determinations of \vsini\ and 
$\lambda$ are degenerate \citep{2011A&A...531A..24T,2011ApJ...738...50A,2011A&A...534A..16A}. 
To ensure \vsini\ is consistent with our spectroscopic measurement, we imposed 
a Gaussian prior on it by means of a Bayesian penalty on 
$\chi^2$, with mean and variance as determined from the HARPS spectra 
(see Section \ref{sec:stars}). 
From an MCMC run with no prior we obtained $\lambda = -8 \pm 54 \degr$ and \vsini$ = 4.1^{+5.9}_{-0.9}$ \kms, 
with \vsini\ reaching values as high as 30\,\kms, which is clearly inconsitent with the spectral measurement of 
\vsini~$= 3.1 \pm 0.6$ \kms. 
The degeneracy between \vsini\ and $\lambda$ could be broken by line-profile 
tomography \citep{2007A&A...474..565A,2010MNRAS.403..151C}.

\begin{figure*}
\centering
\includegraphics[width=168mm]{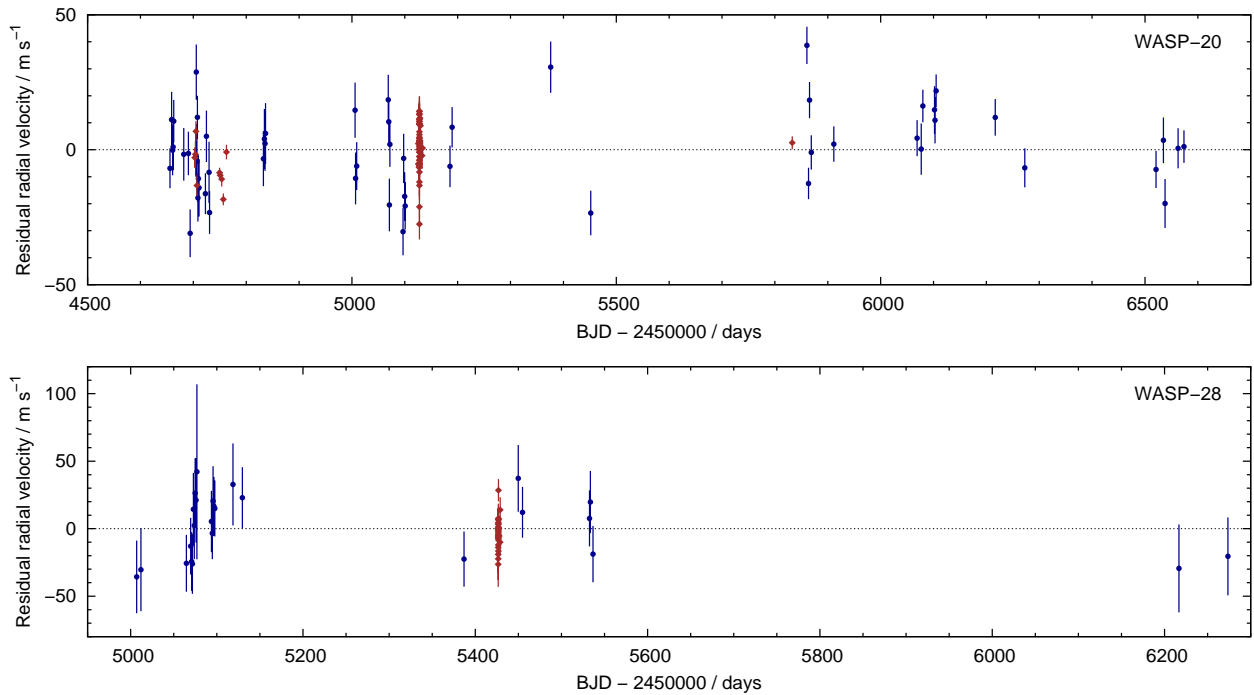}
\caption{The residuals about the best-fitting Keplerian orbits as a function of time 
(CORALIE in blue, HARPS in brown). 
There is no evidence of additional bodies. \label{fig:resid}}
\end{figure*}

The median values and the 1-$\sigma$ limits of our MCMC parameters' posterior 
distributions are given in 
Table~\ref{tab:mcmc} along with those of the derived parameters. 
The best fits to the radial velocities and the 
photometry are plotted in Figures~\ref{fig:w20-rv-phot} and 
\ref{fig:w28-rv-phot}, with the RM effects highlighted in Figure~\ref{fig:rm}. 
There is no evidence of additional bodies in the RV residuals 
(Figure~\ref{fig:resid}).

We checked the effect of splitting the HARPS RVs for WASP-20 into two sets: those taken 
around the spectroscopic transit observation, spanning 2009 Oct 20--28, and the remainder. 
The fit to the RM effect was very similar to that when treating the HARPS RVs as a single dataset 
($\lambda = 11.9 \pm 3.8 \degr$; \vsini\ = $4.75 \pm 0.51$ \kms).

\begin{figure}
\centering
\includegraphics[width=84mm]{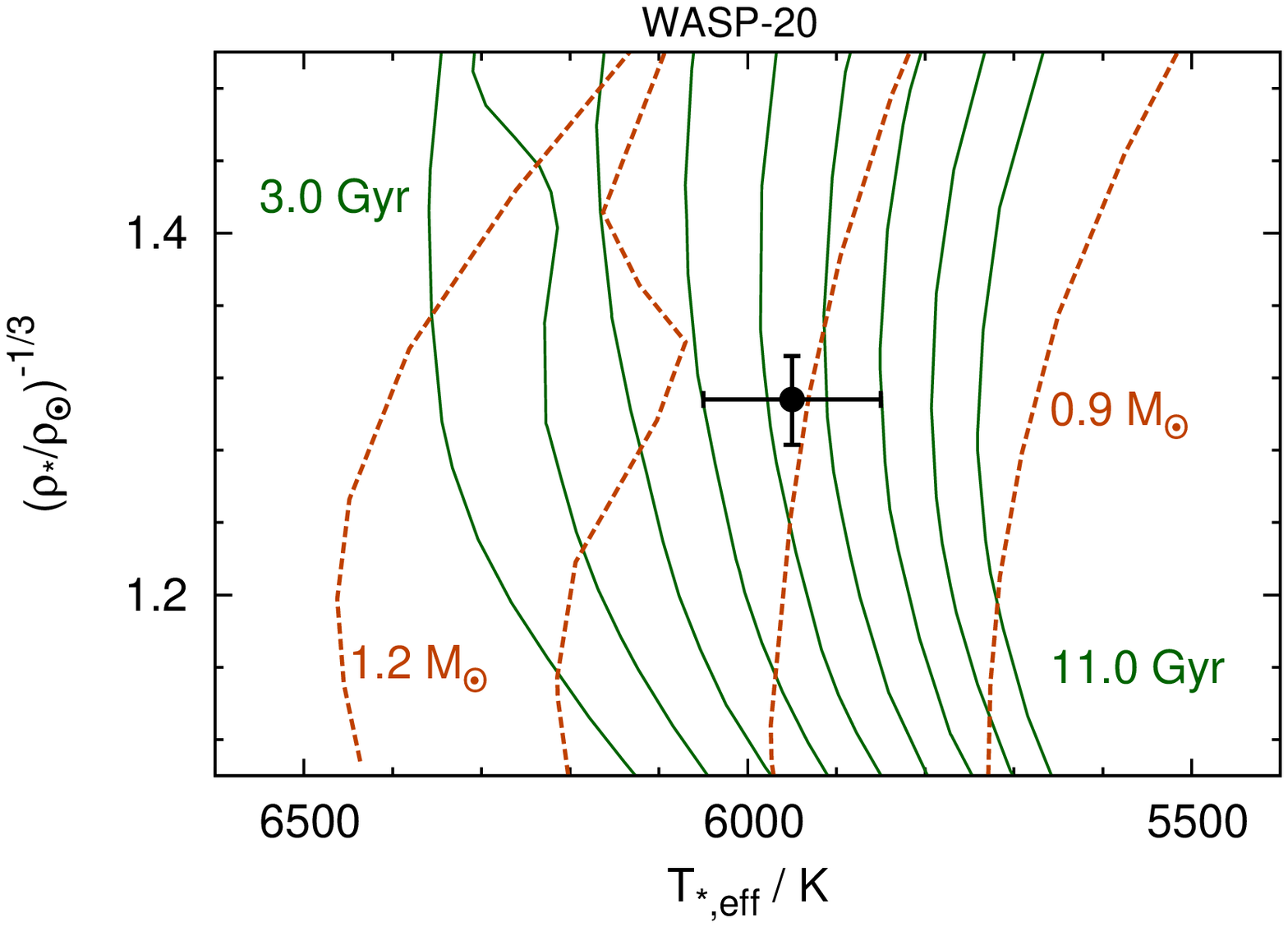}
\includegraphics[width=84mm]{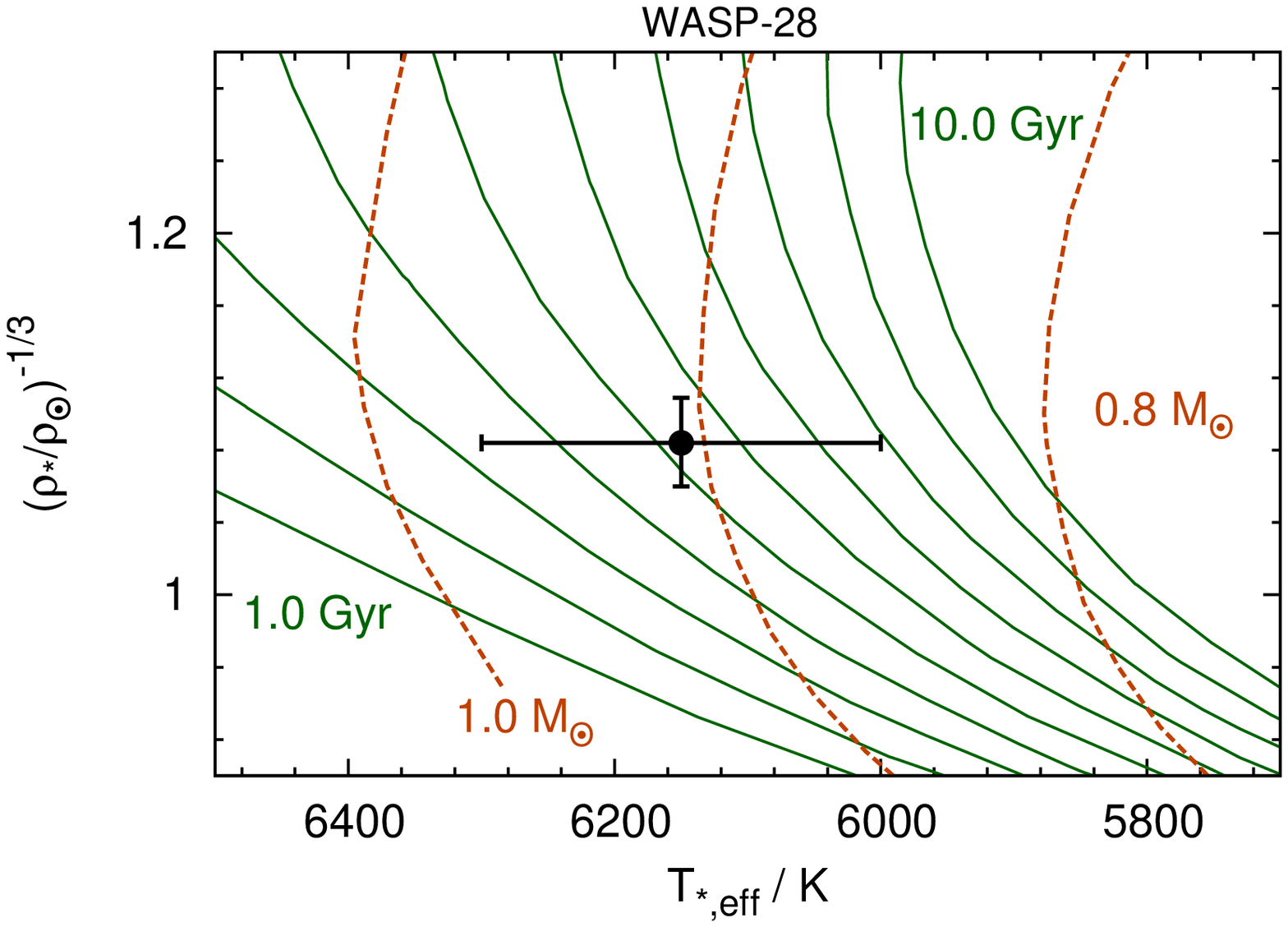}
\caption{Modified H--R diagrams. The isochrones, from \citet{2012MNRAS.427..127B}, 
are in steps of integer Gyr with $Z=0.0152$ for WASP-20 and $Z=0.0078$ for WASP-28. 
The mass tracks, from \citet{2008A&A...484..815B}, are in steps of 0.1\,\msol\ with $Z=0.019$ for WASP-20 and $Z=0.0097$ for WASP-28. 
\label{fig:evol}}
\end{figure}

We interpolated the mass tracks of \citet{2008A&A...484..815B} and the isochrones of \citet{2012MNRAS.427..127B} using \densstar, \teff\ and \feh\ from the MCMC analysis (Figure~\ref{fig:evol}). 
This suggests an age of $7^{+2}_{-1}$\,Gyr for WASP-20 and $5^{+3}_{-2}$\,Gyr for 
WASP-28.
The mass tracks suggest a stellar mass of $1.02^{+0.07}_{-0.03}$\,\msol\ for WASP-20 
and $0.91 \pm 0.06$\,\msol\ for WASP-28. 
These are slightly lower than the masses derived from our MCMC analyses 
($1.20 \pm 0.04$\,\msol\ for WASP-20 and $1.02 \pm 0.05$\,\msol\ for WASP-28), 
though they are consistent at the 2.3-$\sigma$ (WASP-20) and 1.4-$\sigma$ (WASP-28) levels. 

\begin{table*} 
\caption{System parameters from the MCMC analyses} 
\label{tab:mcmc}
\begin{tabular}{lcccc}
\hline
Parameter & Symbol & WASP-20 & WASP-28 & Unit\\ 
\hline 
\\
Orbital period & $P$ & $4.8996285 \pm 0.0000034$ & $3.4088300 \pm 0.000006$ & d\\
Epoch of mid-transit & $T_{\rm c}$ & $2\,455\,715.65562 \pm 0.00028$ & $2\,455\,290.40519 \pm 0.00031$ & BJD (UTC)\\
Transit duration & $T_{\rm 14}$ & $0.1416 \pm 0.0013$ & $0.1349 \pm 0.0010$ & d\\
Transit ingress/egress duration & $T_{\rm 12}=T_{\rm 34}$ & $0.0263 \pm 0.0016$ & $0.01441 \pm 0.00070$ & d\\
Scaled orbital major semi-axis & $a$/\rstar & $9.29 \pm 0.23$ & $8.79 \pm 0.19$ & \\
Planet-to-star area ratio & $\Delta F=R_{\rm P}^{2}$/R$_{*}^{2}$ & $0.01161 \pm 0.00022$ & $0.01300 \pm 0.00027$ & \\
Impact parameter & $b$ & $0.718 \pm 0.018$ & $0.21 \pm 0.10$ & \\
Orbital inclination & $i_{\rm P}$ \medskip & $85.57 \pm 0.22$ & $88.61 \pm 0.67$ & \deg\\
Stellar reflex velocity semi-amplitude & $K_{\rm 1}$ & $33.0 \pm 1.7$ & $120.5 \pm 4.2$ & \ms\\
Systemic velocity & $\gamma$ & $1\,316.222 \pm 0.080$ & $24\,216.41 \pm 0.43$ & \ms\\
Offset between HARPS \& CORALIE & $\Delta \gamma_{\rm HARPS}$ & $16.10 \pm 0.74$ & $6.0 \pm 1.2$ & \ms \\
Eccentricity & $e$  & 0 (adopted; $<$0.11 at 2\,$\sigma$) & 0 (adopted; $<$0.14 at 2\,$\sigma$) & \\
Sky-projected spin-orbit angle & $\lambda$ &  $8.1 \pm 3.6$ & $8 \pm 18$ & \deg\\
Sky-projected stellar rotation velocity \medskip & \vsini & $4.71 \pm 0.50$ & $3.25 \pm 0.34$ & \kms\\
Stellar mass & \mstar &  $1.202 \pm 0.040$ & $1.021 \pm 0.050$ & \msol\\
Stellar radius & \rstar & $1.392 \pm 0.044$ & $1.094 \pm 0.031$ & \rsol\\
Stellar surface gravity & $\log g_{*}$ & $4.232 \pm 0.020$ & $4.370 \pm 0.018$ & (cgs)\\
Stellar density & \densstar & $0.447 \pm 0.033$ & $0.784 \pm 0.058$ & \denssol\\
Stellar effective temperature & $T_{\rm eff}$ & $5950 \pm 100$ & $6150 \pm 140$ & K\\
Stellar metallicity & {[Fe/H]} \medskip & $-0.009 \pm 0.061$ & $-0.290 \pm 0.10$ & \\
Planetary mass & \mplanet & $0.313 \pm 0.018$ & $0.907 \pm 0.043$ & \mjup\\
Planetary radius & \rplanet & $1.458 \pm 0.057$ & $1.213 \pm 0.042$ & \rjup\\
Planetary surface gravity & $\log g_{\rm P}$ & $2.527 \pm 0.036$ & $3.149 \pm 0.028$ & (cgs)\\
Planetary density & \densplanet & 0.1006$^{+ 0.0131}_{- 0.0099}$& $0.508 \pm 0.047$ & \densjup\\
Orbital major semi-axis & $a$ & $0.06003 \pm 0.00067$ & $0.04469 \pm 0.00076$ & AU\\
Planetary equilibrium temperature & $T_{\rm eql}$ & $1379 \pm 32$ & $1468 \pm 37$ & K\\
\\ 
\hline 
\end{tabular} 
\end{table*} 

\section{Discussion}
We present the discovery of WASP-20b, a Saturn-mass planet in a 4.9-day 
orbit around CD-24~102, and WASP-28b, a Jupiter-mass planet in a 3.4-day orbit 
around a $V$=12 star. 
Based on their masses, orbital distances, irradiation levels and metallicities, 
the radii of the planets (1.46\,\rjup\ for WASP-20b and 1.21\,\rjup\ for WASP-28b) 
are consistent with the predictions of the empirical relations of 
\citet{2012A&A...540A..99E}: 1.33\,\rjup\ for WASP-20b and 1.26\,\rjup\ for WASP-28b. 

We find both planets to be in prograde orbits and, as the \vsini\ values are consistent 
with expectations for $v$ for such stars, it is probable that $I_*\approx 90\degr$ and so $\Psi \approx \lambda$.
Assuming the spin axis of WASP-20 to be in the sky plane, the orbit of WASP-20b 
is aligned or near-aligned with the stellar spin ($\lambda = 8.1 \pm 3.6 ^\circ$). 
Although our determination for WASP-28b ($\lambda = 8 \pm 18 ^\circ$) is 
consistent with an aligned orbit, its low impact parameter results in a greater 
uncertainty. 

Both host stars are near the posited boundary between efficient and inefficient 
aligners of \teff\ $\approx$ 6250\,K \citep{2010ApJ...718L.145W,2010ApJ...719..602S}.
Both systems, with their low obliquity determinations, are consistent with the observation that 
systems with cool host stars and short expected tidal timescales are aligned (\citealt{2012ApJ...757...18A} and references therein).
With \mplanet/\mstar\ = 0.00025 and $a$/\rstar\ = 9.3, WASP-20b experiences relatively 
weak tidal forces. Its relative tidal dissipation timescale of $\tau_{\rm CE} = 2.6 \times 10^{13}$ yr 
is the longest for the cool-host group barring HD\,17156\,b (see section 5.3 of 
\citealt{2012ApJ...757...18A}). 
\footnote{Note that these timescales are relative and that \citet{2012ApJ...757...18A} plot 
the timescales divided by $5 \times 10^9$.}
With \mplanet/\mstar\ = 0.00084 and $a$/\rstar\ = 9.5, $\tau_{\rm CE} = 2.5 \times 10^{12}$ yr for WASP-28b, placing it within the main grouping of aligned, cool-host systems. 
Assuming the planets did not arrive in their current orbits recently, the advanced 
age of both systems, $7^{+2}_{-1}$ Gyr for WASP-20 and $5^{+3}_{-2}$ Gyr for WASP-28, 
mean tidal dissipation would have occurred over a long timescale. 

The planets may have arrived in their current orbits via planet-disc migration \citep{1996Natur.380..606L,2009ApJ...705.1575M} or via the scattering and 
circularisation route 
\citep{2007ApJ...669.1298F,2008ApJ...678..498N,2010ApJ...725.1995M,
2011Natur.473..187N}.
Assuming the latter, the planets may have scattered into aligned orbits or they may 
have aligned with the spins of their host stars via tidal interaction, though how 
this could have occurred without the planets falling into their stars is currently a 
mystery \citep{2009MNRAS.395.2268B}. 

\section*{Acknowledgements}
WASP-South is hosted by the South African Astronomical Observatory and SuperWASP-North is hosted by the Isaac Newton Group on La Palma. We are grateful for their ongoing support and assistance. 
Funding for WASP comes from consortium universities and from the UK's Science and Technology Facilities Council. 
M. Gillon is a FNRS Research Associate. 
A. H. M. J. Triaud is a Swiss National Science Foundation fellow under grant number PBGEP2-145594. 


\end{document}